\documentclass[sigconf]{acmart}
\AtBeginDocument{%
  }

\setcopyright{acmlicensed}
\copyrightyear{2018}
\acmYear{2018}
\acmDOI{XXXXXXX.XXXXXXX}
\acmConference[Conference acronym 'XX]{Make sure to enter the correct
  conference title from your rights confirmation email}{June 03--05,
  2018}{Woodstock, NY}
\acmISBN{978-1-4503-XXXX-X/2018/06}

\usepackage{enumitem}

\begin{document}

\title{LongRetriever: Towards Ultra-Long Sequence based Candidate Retrieval for Recommendation}

\author{Qin Ren}
\email{renqin.97@bytedance.com}
\affiliation{%
  \institution{ByteDance}
  \city{Beijing}
  \country{China}
}

\author{Zheng Chai\textsuperscript{†}}
\thanks{\textsuperscript{†}Corresponding Authors.}
\email{chaizheng.cz@bytedance.com}
\affiliation{%
  \institution{ByteDance}
  \city{Hangzhou}
  \country{China}
}

\author{Xijun Xiao}
\email{xiaoxijun@bytedance.com}
\affiliation{%
  \institution{ByteDance}
  \city{Beijing}
  \country{China}
}

\author{Yuchao Zheng\textsuperscript{†}}
\email{zhengyuchao.yc@bytedance.com}
\affiliation{%
  \institution{ByteDance}
  \city{Beijing}
  \country{China}
}

\author{Di Wu}
\email{di.wu@bytedance.com}
\affiliation{%
  \institution{ByteDance}
  \city{Beijing}
  \country{China}
}

\renewcommand{\shortauthors}{Trovato et al.}

\begin{abstract}
Precisely modeling user ultra-long sequences is critical for industrial recommender systems. Current approaches predominantly focus on leveraging ultra-long sequences in the ranking stage, whereas research for the candidate retrieval stage remains under-explored. This paper presents LongRetriever, a practical framework for incorporating ultra-long sequences into the retrieval stage of recommenders. Specifically, we propose in-context training and multi-context retrieval, which enable candidate-specific interaction between user sequence and candidate item, and ensure training-serving consistency under the search-based paradigm. Extensive online A/B testing conducted on a large-scale e-commerce platform demonstrates statistically significant improvements, confirming the framework's effectiveness. Currently, LongRetriever has been fully deployed in the platform, impacting billions of users.


\end{abstract}




\maketitle


\section{INTRODUCTION}
Retrieving highly relevant items from a massive item repository within constrained timeframes presents a significant challenge in industrial recommendation systems\cite{roy2022systematic, bondevik2024systematic}. The predominant solution employs a dual-stage methodology \cite{covington2016deep, hron2021component}, wherein an initial \textit{candidate retrieval} stage is dedicated to narrowing down the candidate set of items, enabling subsequent deployment of more precise yet computationally intensive models during the \textit{ranking} stage. 

Generally, popularity-based filtering and collaborative filtering have been widely adopted in candidate retrieval stage. With the advancement of deep learning, embedding-based retrieval (EBR) has emerged as the predominant paradigm \cite{huang2020embedding, li2021embedding}. EBR utilizes a dual-tower deep neural architecture to encode user and item features into dense vector representations, respectively. It optimizes a distance metric (e.g., cosine similarity) between these vectors within a discriminative learning framework, distinguishing relevant items (positives) from irrelevant ones (negatives). During deployment, item vectors are precomputed and indexed using approximate nearest neighbor (ANN) search techniques \cite{aumuller2020ann}(e.g., HNSW\cite{malkov2018efficient}), enabling efficient top-k item retrieval in sub-linear time for online serving.

Despite a favorable trade-off between efficiency and efficacy, constrained by computational resources and online latency requirements, EBR typically processes only short user sequences (e.g., the most recent 200 behaviors) in the User Encoder. Besides, due to the separated dual-tower learning paradigm, it is generally difficult for the candidate item to interact with user-side historical behavioral sequences, leveraging a remarkable margin for understanding user interest deeply. Precisely modeling user sequences is essential to current recommenders, in which typical approaches generally adopt a target-attention mechanism for modeling user interest, and the most representative approaches include DIN \cite{zhou2018deep}, DIEN \cite{zhou2019deep}, CAN \cite{bian2022can}, etc. With the advancement of computational capabilities and the ever-growing accumulation of user online behaviors, recent research has shifted toward achieving more comprehensive modeling of ultra-long sequences in recommender systems \cite{pi2020search, chai2025longer}. 

Despite the popularity, it is noted that current long-sequence modeling is mostly developed for the ranking stage in recommenders. How to leverage the power of ultra-long sequence for enhancing EBR remains a notable gap in industrial recommenders. Generally, there are the following challenges in leveraging ultra-long sequence in candidate retrieval:

\begin{figure*}[t!]
  \centering
  \includegraphics[width=\linewidth]{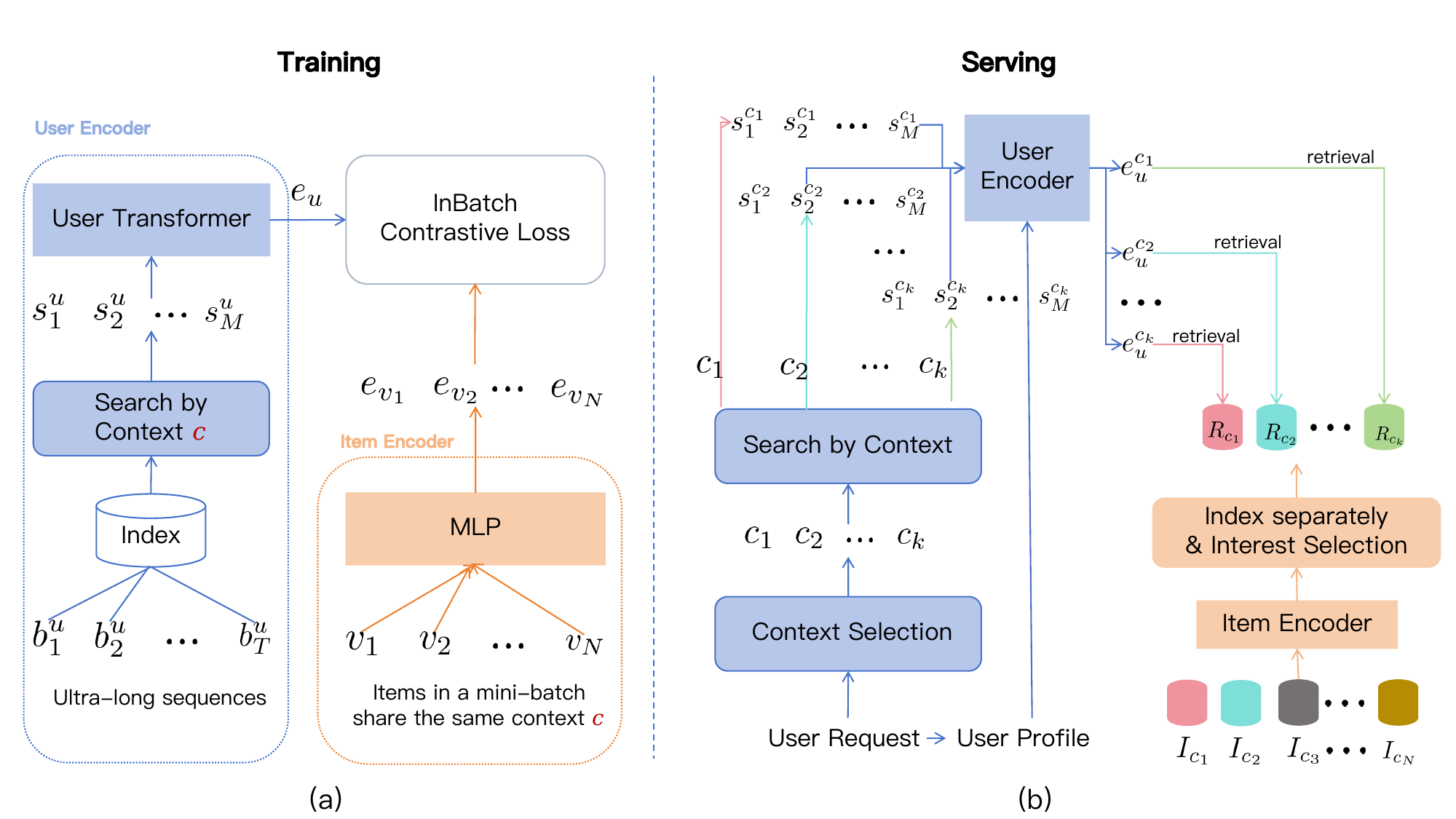}
  \caption{LongRetriever Model Framework.}
\end{figure*}



\begin{itemize}[leftmargin=*]
\item Adequate interaction between the ultra-long sequence and candidate item. Note that the popular target attention-based approaches are applicable in ranking stage due to the limited number of candidate items (around hundreds in industry). While in the matching stage, the candidate items scale to the order of hundreds of billions, making the interaction between the ultra-long sequence and candidate item intractable in practice.
\item Diverse while precise interest modeling with controllable learning capability with ultra-long sequences. Traditional dual-tower paradigms generally encode user representation as a single vector which bottlenecks modeling diverse interests at industrial scale, and existing multi-interest modules \cite{li2019multi, chai2022user} generally lack concrete semantics, leading to reduced controllability. 
\end{itemize}

To tackle the above challenges, in this paper, we present \textit{LongRetriever}, a controllable and accurate diverse-interests learning framework for candidate matching with ultra-long sequences. LongRetriever defines mutually exclusive item categories as interpretable user interests and conducts independent candidate training for each interest under shared model parameters.  It employs a search-based mechanism to filter relevant subsequences from the user’s lifelong behavior sequence for each interest. This approach represents users as multiple dense vectors across different interest domains without increasing training overhead. The utilization of the lifelong user behavior sequence, which demonstrates strong effectiveness in the ranking stage, also enables LongRetriever to achieve more accurate expression within each interest in the candidate matching stage. Overall, the contributions are mainly summarized as follows:
\begin{itemize}
\item We devised LongRetriever, a controllable multi-interest learning framework based on long sequence. To the best of our knowledge, it makes the first attempt to introduce ultra-long sequence to the candidate matching stage.
\item We designed novel in-context training and multi-context serving for introducing user-candidate interaction and ensuring training-serving consistency under the search-based paradigm. Combined with interest selection, this renders the entire retrieval fully interpretable and controllable.
\item Extensive online A/B testing on a super e-commerce platform with billions of users validates LongRetriever's effectiveness. The framework is now fully deployed at industrial scale, impacting billions of users.
\end{itemize}


\section{METHODOLOGY}

\subsection{Background: Embedding-based Retrieval}
Let $U$ and $I$ denote the user and item sets respectively. Given a user $u \in U$ with raw behavior sequence $B_u = [b^u_1, b^u_2, \cdots,b^u_T]$, user basic features $P_u$ including user profiles and context features, and a target item $v\in I$. Embedding-based retrieval frameworks employ a user encoder $f$ and an item encoder $g$ to project user and item features into low-dimensional dense vectors: $e_u = f(B_u, P_u)$ and $e_v = g(v)$. The relevance of user $u$ to item $v$, denoted by $r_{uv}$, is quantified using a distance metric (typically cosine similarity) between $x_u$ and $x_v$. During training, the relevance scores $r_{uv}$ are optimized via an in-batch contrastive loss to distinguish between items interacted by $u$ and non-interacted items within the same mini-batch. During deployment, all item vectors $\{e_v\} \, v \in I$ are precomputed and indexed using approximate nearest neighbor search techniques. Consequently, only the user vector $e_u$ requires real-time computation, and the ANN system can retrieve the k nearest items to $e_u$ from the item repository in sub-linear time.

\subsection{Overall Framework}
Figure 1 illustrates the overall architecture of our LongRetriever model. Generally, given the list of user lifelong behaviors $B_u = [b^u_1, b^u_2, \cdots,b^u_T]$
 where $T$ is the length of user behaviors. LongRetriever first employs category matching to filter the most recent L behaviors from $B_u$ that exactly match the candidate item’s \textit{context}, similar to the Hard-Search method in SIM\cite{pi2020search}. Here, the \textit{context} can be any criteria that can be relied on in practice. In this paper, we adopt item's \textit{category} for hard search. After search based on context, the new user subsequence is denoted as $S_u = [s^u_1, s^u_2, \cdots,s^u_M]$.
 
 LongRetriever subsequently employs a user transformer model as the user encoder $f$ to extract the user vector $e_u$. The input tokens to this Transformer comprise:  
- A $[CLS]$ token to aggregate information across all tokens. 
- $K$ tokens derived from user profile features $P_u = [p^u_1, p^u_2, \cdots,p^u_K]$ to enhance fine-grained interactions between the behavior sequence and basic user features.
-  $L$ filtered sub-behaviors $S_u$.
After adding learnable positional encodings and token type embeddings to the input tokens, LongRetriever utilizes a standard Pre-Layer Normalization Transformer to model the tokens. The final representation of the $[CLS]$ token is adopted as the user representation vector $e_u$:
\begin{equation}
e_u = Transformer([CLS, p^u_1, p^u_2, \cdots,p^u_K, s^u_1, s^u_2, \cdots,s^u_M])
\end{equation}
It is noteworthy that EBR's training requires independent computation of user and item encoders, to meet the requirements of the ANN paradigm during deployment. Consequently, candidate item features are excluded from our User Transformer.

Based on the model structure design, there are two key components designed in LongRetriever, In-Context Training and Multi-Context Retrieval.

\subsection{In-Context Training}
Directly leveraging the search-based long sequence yields the following loss function:

\begin{equation}
\mathcal{L}\left(e_u\right)=-\log \left(\frac{e^{e_u^{\rm T} e_v
}} {e^{e_u^{\rm T} e_v}+\sum\limits_{{v'} \in \mathcal{B}} e^{e_u^{\rm T} e_{v'}}}\right)
\end{equation}

\noindent where the $\mathcal{B}$ denotes the in-batch negative item set, the $v$ denotes the positive item and the $v'$ indicates an arbitrary negative item.

While leveraging the lifelong behavior sequence resulted in significant improvements in offline evaluation metrics, this improvement is proved unreliable, because $S_u$ in $e_u$ implicitly incorporates partial information about the candidate item. This introduces data leakage under the in-batch contrastive objective, where instances influence each other. Consider an instance of interaction between user $u$ and item $v$, where the context of $S_u$ and $v$, i.e., the category of the items, is "Food." This $(u,v)$ pair is treated as a positive example, whereas pairs of $u$ with items from other instances in the current mini-batch (which may belong to categories like "Clothes" or "Pets") are treated as negative examples. Under these conditions, the model can readily reduce the in-batch contrastive loss by leveraging whether the category of $S_u$ matches that of $v$, thereby artificially inflating offline metrics.



To mitigate this data leakage, LongRetriever adopts an In-Context Training strategy. In-Context Training fundamentally involves restructuring samples within each mini-batch, using caching mechanisms to ensure that the subsequence $S_u$ and the candidate item $v$ within the same mini-batch share only one interest category.

Before In-Context Training, EBR’s objective equates to discriminating whether a user would purchase items in the "Food" category over the entire sample space, given the category-specific sequence $S_u$. After In-Context Training, EBR’s learning objective changes to discriminating whether a user would purchase a candidate item over the entire "Food" category sample space, given the category-specific sequence $S_u$. 


\subsection{Multi-Context Retrieval}
The In-Context Training enables user-candidate interaction during training. While in serving stage, due to the separation inference of the user and item tower, it is of significance to infer several categories that users may be interested in to enable multi-context retrieval based on user's ultra-long behavior sequence. Here, two necessary stages are devised for user interest estimation and multi-interest retrieval.

\begin{table*}[ht]
  \caption{Online A/B Test Results on Business Metrics}
  \label{tab:freq}
  \begin{tabular}{cccccl}
    \toprule
    PV & UV CTR & UV CVR & Orders Per User & Exposed Categories Per User & Clicked Categories Per User\\
    \midrule
    \textbf{+0.62\%} & +0.17\% & \textbf{+1.33\%} & \textbf{+1.70\%} & \textbf{+0.14\%} & \textbf{+1.39\%}\\
  \bottomrule
\end{tabular}
\end{table*}

\noindent \textbf{Interest Selection and Representations.} LongRetriever selects specific interests based on user historical behaviors and executes the user encoder multiple times to generate distinct user representations for each interest. Due to computational constraints, the number of interests selected per request should not be excessive. To this, we provide a simple automated interest selection strategy, \textit{Random in Top}. It computes a user’s engagement score for each interest $c_i$ by aggregating historical behavioral data across interests using time-weighted summation:
\begin{equation}
Score_{c_i} = \sum_{b_i^u \in B_u} \frac{sign(c_i = c_{b_i^u})}{t_{b_i^u}}
\end{equation}
where $c_{b_i^u}$ denotes the interest category associated with behavior $b_i$, and $t_{b_i^u}$ represents the time difference (in days) between the occurrence timestamp of behavior $b_i$ and the current request timestamp. \textit{Random in Top} first selects the $top-M$ interests by engagement score from the entire interest set, then randomly selects $N$ interests from this subset for each user request.

\noindent \textbf{Multi-Interest Retrieval.} Following In-Context Training, the user vector is trained exclusively to mine relevant items within a specific interest domain. Consequently, LongRetriever partitions the entire item repository into multiple independent sub-repositories based on category and constructs indexes for each sub-repository. Upon receiving a user request, LongRetriever executes multiple retrievals across sub-repositories associated with selected interests and merges the results using predefined strategies.

\section{EXPERIMENT}
\subsection{Experimental Setting}


We evaluate LongRetriever in our shopping mall, a real-world large-scale industrial recommendation scenario. Unlike the targeted ranking stage, industrial matching systems are highly complex, typically comprising dozens of distinct retrieval strategies. Consequently, offline metrics alone are insufficient for evaluating retrieval effectiveness. We first train models on industrial datasets, then conduct a 7-day online A/B test with real users for evaluation. The training set comprises user interaction logs from our shopping mall, covering 900 million users, 150 million items, and over 10 billion samples spanning 400 consecutive days. Each sample includes user demographic features (e.g., user ID, gender), candidate item, a 50-behavior subsequence filtered from the user's raw lifelong behavior sequence—which averages over 20,000 interactions—based on the candidate item's category, and a label indicating whether there has been an interaction or not. We benchmark LongRetriever against MIND\cite{li2019multi} — a widely adopted multi-interest retrieval strategy deployed in our production environment. LongRetriever adopts the Random 5 in Top 20 strategy to obtain 5 interests.

\begin{table}[ht]
  \caption{Online A/B Test Results On Intermediate Metrics}
  \label{tab:freq}
  \begin{tabular}{cccl}
    \toprule
    Model & AER & UER & CTR*CVR \\
    \midrule
    MIND & \textbf{31.82\%} & 4.58\% & 6.89\% \\
    LongRetriever & 18.72\% & \textbf{9.29\%} & \textbf{7.92\%} \\
  \bottomrule
\end{tabular}
\end{table}

\begin{table}[ht]
  \caption{Ablation Study on Random Interests}
  \label{tab:freq}
  \begin{tabular}{cccl}
    \toprule
    Model & AER & UER & CTR*CVR \\
    \midrule
    Random5 in Top20 & 18.72\% & \textbf{9.29\%} & 7.92\% \\
    Top5 & \textbf{30.43\%} & 7.23\% & \textbf{8.29\%} \\
  \bottomrule
\end{tabular}
\end{table}

\begin{table}[ht]
  \caption{Ablation Study on Lifelong Behavior Sequence}
  \label{tab:freq}
  \begin{tabular}{cccl}
    \toprule
    Model & AER & UER & CTR*CVR \\
    \midrule
    LongRetriever & \textbf{18.72\%} & \textbf{9.29\%} & \textbf{7.92\%} \\
    w/o Long Sequence & 11.45\% & 5.92\% & 5.33\% \\
  \bottomrule
\end{tabular}
\end{table}

\subsection{Experimental results}

Table 1 summarizes the performance of LongRetriever compared to MIND in our shopping mall. Due to commercial policies, we report only relative increases for business metrics and bold items with statistically significant results ($p<0.05$). In terms of convert efficiency, LongRetriever improved item page views (PV) by 0.62\%, unique visitor (UV) click-through rate (CTR) by 0.17\%, and UV conversion rate (CVR) by 1.33\% compared to MIND. This collectively resulted in a 1.7\% increase in average orders per user. Regarding ecosystem development, LongRetriever enhanced recommendation diversity by increasing average exposed categories per user by 0.14\% and average clicked categories per user by 1.39\%.

Table 2 presents intermediate metrics of retrieved items, revealing potential reasons for the business metric improvements in Table 1. All exposure ratio (AER) is used to measure the preference degree of a recommender for all results retrieved by the model. Unique exposure ratio (UER), on the other hand, is used to measure the preference degree of the recommendation system for items that can only be retrieved by this model. LongRetriever exhibits a significantly lower AER than MIND, primarily because MIND retrieves from the full item repository per user vector, whereas LongRetriever uses only 5 categories for Multi-Context Retrieval, severely limiting its candidate scope. However, LongRetriever achieves a 4.71\% higher UER compared to MIND. This indicates that LongRetriever retrieves items highly distinct from the items retrieved by other existing strategies. Furthermore, the unique conversion efficiency of LongRetriever, measured by CTR * CVR, improves by 1.03\% over MIND, thus contributing to business metric gains.

\subsection{Ablation study}

\noindent \textbf{Random Interests.} Table 3 compares two different interest selection strategies. User interest distributions typically exhibit long-tailed characteristics. When retaining only the top 5 interests in interest selection, the AER increased by 11.71\% compared to the Random in Top strategy. Additionally, the conversion efficiency improved by 0.37\%. However, the UER is decreased by 2.06\%. Consequently, the overall performance was inferior to Random in Top. Furthermore, Random in Top's inherent explicit diversity structure contributes positively to ecosystem development.

\noindent \textbf{Lifelong behavior sequence.} Table 4 conducted an ablation study on the impact of lifelong behavior sequence while retaining In-Context training and multiple In-Context retrieval. Without the aid of effective lifelong behavior sequence, LongRetriever struggled to effectively complete the training paradigm where all negative examples were hard negatives in In-Context training, resulting in a significant decline across all posterior metrics.

\noindent \textbf{In-Context training and In-Context retrieval.} Table 5 ablates the impact of In-Context training and In-Context retrieval while retaining the lifelong behavior sequence. Removing the In-Context components introduces data leakage in LongRetriever's training. Although the absence of these components yields a significant increase in all exposure ratio and unique exposure ratio, the retrieved candidates exhibit mediocre performance in conversion efficiency.

\begin{table}[ht]
  \caption{Ablation Study on In-Context Components}
  \label{tab:freq}
  \begin{tabular}{cccl}
    \toprule
    Model & AER & UER & CTR*CVR \\
    \midrule
    LongRetriever & 18.72\% & 9.29\% & \textbf{7.92\%} \\
    w/o In-Context & \textbf{38.92\%} & \textbf{13.48\%} & 5.61\% \\
  \bottomrule
\end{tabular}
\end{table}


\section{CONCLUSIONS}
In this paper, LongRetriever is proposed to address the challenge of modeling more controllable and accurate diverse user interests in real industry. LongRetriever employs independent In-Context Training and Multi-Context Retrieval for each interest. LongRetriever integrates the powerful user's lifelong behavior sequence using a search-based approach. This transforms the single fixed-length user vector into multiple precise vectors within their respective interest domains, with minimal additional training overhead. Extensive online A/B testing within our business scenarios validates LongRetriever’s effectiveness. The framework is now deployed at scale, impacting billions of users.

In future work, user information could be processed by large language models to generate optimized category assignments, thereby achieving more accurate and diverse recommendations.


\balance
\bibliographystyle{ACM-Reference-Format}
\bibliography{ref}

\end{document}